# The Importance of Computing Education Research

Steve Cooper, Jeff Forbes, Armando Fox, Susanne Hambrusch,

Andrew Ko, and Beth Simon

Version 2

January 14, 2016

## 1. Introduction

Interest in computer science is growing. As a result, computer science (CS) and related departments are experiencing an explosive increase in undergraduate enrollments and unprecedented demand from other disciplines for learning computing. According to the 2014 CRA Taulbee Survey, the number of undergraduates declaring a computing major at Ph.D. granting departments in the US has increased 60% from 2011-2014 and the number of degrees granted has increased by 34% from 2008-2013.

However, this growth is not limited to higher education. New York City, San Francisco and Oakland public schools will soon be offering computer science to all students at all schools from preschool to 12th grade, although it will be an elective for high school students. Chicago has gone even further, pledging to make a yearlong computer science course a high school graduation requirement. Tens of thousands of adults are paying to attend for-profit "developer boot camps". Hundreds of thousands of learners are enrolling in MOOCs (Massive Open Online Courses) on computing and millions more are learning with online tutorials, such as those provided by *Code.org* and *Codecademy.org*. From 2013 to 2014, this increase in demand led to an increase of 26% in the number of high school students taking the AP CS A exam [22], with even more are expected to take the redesigned AP CS Principles course that launches this year. This unprecedented demand means that CS departments are likely to teach not only *more* students in the coming decades, but *more diverse* students, with more varied backgrounds, motivations, preparations, and abilities.

This growth is an unparalleled opportunity to expand the reach of computing education. However, this growth is also a unique research challenge, as we know very little about how best to teach our current students, let alone the students soon to arrive. The burgeoning field of **Computing Education Research (CER)** is positioned to address this challenge by answering research questions such as:

- How should we teach computer science, from programming to advanced principles, to a broader and more diverse audience?
- How can we ensure that we retain this more diverse audience through inclusive pedagogy and generally more effective teaching?
- How can teaching approaches and their assessment (regarding student learning) scale effectively?

- What training should K-12 teachers receive? What methods have been shown to be effective?
- How can computer science teaching adapt to how different people learn and build on age related learning progressions?
- How should computing be taught and integrated into other disciplines?

We argue that computer science departments should lead the way in establishing CER as a foundational research area of computer science, discovering the best ways to teach CS, and inventing the best technologies with which to teach it. This is not only in the best long-term interest of our field, but also the long-term interests of society. This white paper provides a snapshot of the current state of CER and makes actionable recommendations for academic leaders to grow CER as a successful research area in their departments.

## 2. CER: Recent Results and Future Opportunities

In this section, we highlight selected research results from CER, providing a glimpse of the transformative impact that further investment in CER will have on computing education and its relationship to computer science research.

Teaching CS to a broader audience

The increased interest in computer science is resulting in a more diverse population of students taking computing courses. This diversity spans many dimensions including type and level of educational background, age, nature of interest in the subject, gender, race, disabilities, and others. This breadth affords both opportunities and responsibilities to teach populations and individuals as effectively as possible. And, our field is uniquely positioned to engage in this effort because some of the educational challenges can leverage and be built upon ideas and solutions from our field.

Research is currently being done at both the undergraduate [1, 13] and at the K-12 [2, 7, 10] levels. For example, Leah Buechley developed Lilpad Arduino to interest young women in computing using e-textiles, and Andreas Stefik at UNLV has explored programming languages accessible to blind and low-vision learners. Research in this space is contributing foundational insights and technologies. We predict that the results will be a significantly broader range of participation in computing than we see today, especially if these efforts build on effective practices developed by NCWIT and others. Expanding the scope of research in this area - as computer science moves towards a permanent role in K-12 education - has the potential for significant impact.

Teaching CS more effectively

Our field needs to improve teacher training, teaching methods, and teaching technologies, and to develop evidence-based methods to measure improvements. Recent work includes: Bjoern Hartmann's lab at Berkeley has invented novel ways to compute hints for DFA constructions that significantly improve problem completion time; Scott Klemmer's lab at UCSD has created Talkabout, a more effective discussion board for peer learning in MOOCs; Chris Piech has developed new machine learning algorithms to model students' progression in solving programming problems [11]. These contributions contain education research as well as compelling computer science research. This includes developing new algorithms, techniques, and systems for leveraging massive data sets, machine learning, and other CS techniques that have the potential to transform other areas of computer science. Other research in better pedagogy includes the use of contextualized learning [8], pair programming [15], peer instruction [12], and flipped classrooms [3].

Teaching CS in all subjects

Computer science is relevant in nearly all disciplines, and yet we know very little about what CS material to teach in each discipline, how to teach it, and what languages, platforms, and technologies to teach it with. Emmanuel Schanzer's Bootstrap [18] teaches algebraic and geometric concepts through programming for middle and early high school students. Irene Lee's Project GUTS [19] incorporates the teaching of computing in middle school science classes. University of Washington Ph.D. student Katerena Kuksenok is investigating the educational role of Greg Wilson's Software Carpentry scientific computing workshops in helping students in geology, biology, chemistry, and physics mine massive oceanographic data sets to inform climate change. New research in this space will forge unprecedented links between computing, engineering, the natural sciences, the social sciences, and the computing education required to train domain experts to fully harness computing in their work.

Teaching CS at scale

A fundamental challenge when teaching to large classes is maintaining the quality of instruction. Scott Klemmer at UCSD has started to investigate scalable small group instruction methods for high enrollment classes. The peer instruction work [12] has done similar investigations. Non-profits such as *Code.org* and *Khan Academy* are also contributing to this space, planning wide-scale deployments and evaluations of new curricula at all grade levels. Andrew Ko and Michael Lee, working with *Gidget* at the University of Washington, have studied online adult education, inventing online learning experiences that require little to no direct instruction to effectively teach introductory concepts, while doing so in half the time of online tutorials. With new investments in computing education research faculty, we will see many more of these efforts creating a vibrant array of learning opportunities that reach a wide diversity of learners across urban, rural, and developing regions at all socioeconomic levels.

Teaching and learning CS at all ages and everywhere

Computer science is increasingly being taught in primary and secondary school—with technologies, curriculum, and teaching methods that do little to account for developmental differences of students of all ages. Researchers such as Diana Franklin, at University of Chicago, and Linda Seiter, at John Carroll University, are exploring learning progressions, determining which subjects can successfully be taught to middle and elementary school students. Others have long investigated ways of scaling instruction to K-12 students, including Mitch Resnick's *Scratch* at MIT, and the late Randy Pausch's *Alice* at Carnegie Mellon. Both have reached hundreds of thousands learners, providing a scalable foundation for teaching computing across the world. New research in this space will help provide CS teachers at all levels with the knowledge, teaching methods, and learning technologies to successfully educate people of all ages.

Computing education researchers are inventing ways for people to learn computing at home, at school, at work, and at play, taking the lessons they learn in one context and applying them in another. Researchers such as Ben Shapiro are embracing this idea, exploring toolkits [6] that allow learners to construct networked devices that interact with the world, empowering students to learn outside of the lab, in their homes and hangouts. New research in this space will contribute new learning technologies that are physical, tangible, offline, and online, ensuring that learning of the big ideas in CS can occur anytime, anywhere.

## 3. Promoting CER in Computer Science Departments

The research highlighted in the previous section is exciting. Many interesting, relevant, and challenging problems remain to be explored and need the expertise from computer science researchers as well as education and learning science researchers. Today, our field is producing a fraction of the discoveries needed to teach computing effectively in universities, colleges, and K-12. To deal with the expected growth and challenges of teaching computer science, we argue that departments should embrace CER as a research area. In general, this requires a culture shift regarding how research on computing education is regarded in the university tenure track. In this section, we discuss some of the core strengths of the computing education research community and some of the key opportunities for maturing CER into a robust, impactful research field.

Faculty

CER is by its very nature interdisciplinary. Research questions in CER often use computer science techniques and approaches, such as machine learning and big data analytics. This relationship has encouraged collaboration between computing education research faculty and traditional CS researchers. However, CER also relies on the social sciences, often leveraging collaborations with education and learning sciences researchers.

Despite the exciting opportunity of meaningful interdisciplinary research, our field has no more than a few dozen computing education researchers in the U.S. Other countries, such as Israel, Finland, and Australia, are more active in computing education research, with greater acceptance of CER as a research discipline in computer science. Despite the opportunities, today there are few academic career pathways in the U.S. for computing education researchers. Faculty appointments are scarce and are often instructors or professors of the practice positions. Some of these appointments have higher teaching loads and/or are primarily expected to teach rather than to conduct research (or support graduate students) in CER. Graduate students interested in CER often don't see it as a viable research field to pursue and interactions with graduate students in education or learning sciences are not common.

Ph.D. students

There are few Ph.D. students in CER. Interested Ph.D. students may fear the lack of tenure-track positions in CER.  Those students who do engage in CER, may feel the need for keeping one leg in CER and one leg in some other widely recruited area  in order to be successful in a tenure-track search. The job outcomes for many CER Ph.D. students have been full-time teaching positions in universities and colleges, or, like many other CS Ph.D. students, lucrative positions in industry having little to do with their research expertise.

Despite these current challenges, the latent potential for participation in computing education research is enormous. We encounter countless undergraduate, MS, and Ph.D. students who are passionate about improving CS education, but just can't find the faculty mentors in computer science departments to help them navigate research opportunities. With strategic investments in this area by CS departments and funding agencies, there is a substantial population of motivated, engaged, and talented students ready to grow and transform the field.

Conferences and Journals

SIGCSE, the ACM Technical Symposium on Computer Science Education, is the largest conference related to CER. Though traditionally dominated by practitioners, SIGCSE has seen increasing participation by researchers; it serves as a key place to directly disseminate discoveries to teachers. ICER, the International Computing Education Research conference, focuses exclusively on research, and typically publishes the most rigorous, evidence-based work in the field. ICER has recently seen rapid growth in participation. The Learning@Scale conference focuses on how learning and teaching can change and improve when engaging large numbers of students; many contributions apply machine learning, NLP, and other techniques to study new approaches for students to learn and for teachers to teach. Strong CER journals have also emerged, with the *ACM Transactions on Computing Education, Computer Science Education, IEEE Transactions on Education*, and *Computers & Education* publishing rigorous, and often more theoretically-grounded foundational work. While these research venues are healthy, they have not yet reached the impact factors for name recognition in the broader CS research community.

Funding

Research funding for CER from government, industry, and foundations has grown in recent years. NSF is funding more basic research in CER thanks to partnerships between CISE and EHR and CISE's commitment towards CS education [17]. Google Research and Microsoft Research have been generous with gifts and grants for basic CER. Other core programs at NSF have also strongly supported work at the intersection of computing education and other areas such as machine learning, PL, and HCI. NSF recently awarded its first CAREER awards for new CER faculty [9], and the NSF Graduate Research Fellowship Program now explicitly supports students pursuing computing education research. Additionally, NSF continues to support research in CER, including broadening participation, effective teacher training, and improving undergraduate CS education with proven methods from basic research. As the community of researchers grows, and the area increases in national importance, we anticipate that these funding resources will also grow.

## 4. Recommendations for Academic Leaders

Creating an environment in which computing education research flourishes and also applies to teaching practice is a long-term endeavor. Public interest in K-12 computing education has increased in recent years and many CS departments have new interests in improving the quality of undergraduate education and student retention, especially retention of members of underrepresented groups through evidence-based practices [25]. The growing public interest, combined with the availability of computing education research funding, creates a unique environment for departments to consider CER as a respected research area.

Although many of the challenges in growing CER are long term, this section presents concrete first steps academic leaders and faculty can take to highlight the relevance and impact of CER, leading to the cultural change needed to make CER a "first-class citizen" in CS departments. We follow the suggestions with a list of other stakeholders who can help effect changes that stretch beyond the boundaries of individual departments or schools.

Faculty

The foundation for growth of CER is to engage faculty and support faculty activities:

- **Connect CS faculty having CER interests**. Many CS departments already have faculty interested in or involved in CER. As they publish in different venues, they may be unaware of each other's work. Faculty should make an effort in finding out the different ways CER related research can happen in CS. Academic leaders are in an especially good position to make CS faculty aware of the work of their colleagues and encourage collaboration.
- **Connect faculty with colleagues in other departments and centers**. Many institutions have departments, centers, or individual faculty in education, learning sciences, or psychology; identifying potential associated researchers in these units is crucial to establishing collaborations and partnerships. These relationships can lead to interdisciplinary courses, joint research and proposals, as well as joint supervision of Ph.D. students interested in computing education research. In addition, many institutions have centers for teaching and learning (CTL). Such centers may currently have little involvement with computing education researchers, but CTLs can play a crucial role in assisting CER faculty on assessment and evaluation, preparing materials, working with other stakeholders, etc.
- **Highlight the impact of CER faculty research**. During annual reviews and eventual promotions, it is necessary to recognize and evaluate the impact of CER as well as to distinguish it from teaching contributions. CS faculty are familiar with researchers getting credit for impacting programming language design, maintaining open source projects, and starting companies. In CER, impact activities might include the wide adoption of a developed tool supporting evidence-based education, or the wide adoption of a curriculum/set of pedagogies for a particular course based on evidence-based practices, or a startup that successfully extends the reach of such tools and practices.
- There exist several models for hiring CER faculty. One approach is for CS Departments to hire CER faculty as regular tenure-track faculty members and to build up CER as a departmental research area. Some departments have found that this approach can attract very strong candidates who identify as computer scientists and best fit into a CS research and teaching environment. The discipline based education research model (DBER) [14], which is already being used to support STEM education researchers in other STEM departments, extends this approach. While situated in the CS Department, DBER researchers are supported in their collaborations with other DBER STEM education researchers in other departments. Alternatively, CER faculty can have joint tenure-track/tenured appointments with other departments (e.g., Education, Learning Sciences, and Cognitive Psychology) with either department serving as the primary department.

Ph.D. students

Parallel to increased CER faculty support and building up CER as a research area, fostering a healthy community of CER doctoral students is important.

- **Signal support for CER online.** Once a department has CER activities, computing education research should be listed as a research area on your departmental website, allowing interested Ph.D. applicants to select it as an interest area.
- **Have flexible doctoral requirements**. CER Ph.D. students need the flexibility to take courses covering social science research methods, learning sciences, and education research, on top of their computing requirements. This can include allowing graduate courses in these areas to count as electives. Departments may want to explore requirements of successful HCI groups as possible models. In addition, breadth exams should allow CER students to go beyond core computing topics.
- **Bootstrap doctoral student funding.** Existing funding opportunities for students in CER are often not utilized. Explicitly encourage CER graduate students to seek NSF Graduate Research Fellowships. Faculty members in areas that potentially interact with CER (HCI, machine learning, PL, software engineering, etc.) should be encouraged to support CER graduate students.

Culture Change

Culture in CS departments may take some time to change. Highly-visible ongoing activities not only support this long-term change, they also send the important signal that CS departments believe in the importance of this agenda. These activities include:

- **Highlight CER speakers.** This can include starting a CER focused lecture series with both CER faculty within the department or outside speakers. Exceptional computing education researchers should be added to a department's Distinguished Lecture Series. CER groups should run informal meetings highlighting recent research results.
- **Support attendance at key CER conferences.** Encourage CER and non-CER faculty to attend ICER, Learning@Scale, or SIGCSE. As these meetings often lead to collaborations and proposal ideas, departments should consider providing travel support as the research areas develops.
- **Present CER discoveries at faculty meetings**. Encourage faculty who do CER research and/or attend CER conferences to give a brief (5 minute) presentation on relevant discoveries to their peers, helping the department to gain awareness of the state of CER.
- **Apply CER to improve departmental courses**. Empower your CER faculty to do research in departmental courses, including courses they do not teach. Offer monetary support or teaching relief for CER faculty who improve your department's teaching.
- **Offer CER fellowships**. Departments could support a "CTLC" (Center for Teaching & Learning in Computing) office within their department, as well as offering prestigious rotating fellowships where a visitor is brought in to research or teach, either as a postdoc or as a visiting fellow.

Other stakeholders

A department doesn't have to do these activities alone, or without resources:

- **Leverage alumni.** Successful, philanthropically-minded CS alumni are often interested in teaching-related activities and novel efforts that improve effective teaching. They could endow postdoc fellowships or provide funds for CER-focused activities.
- **Leverage foundations.** Many foundations (e.g., Spencer, Gates, Hewlett, MacArthur) support CER; submit proposals to provide startup packages/subsidies for new CER faculty hires.
- **Leverage rainmakers.** Many CS departments have faculty who are particularly good at acquiring institutional, state, and private resources. Provide incentives for them to help support and grow CER in your department.
- **Leverage industry.** There are several IT companies with strong presences in education (particularly Google, Intel, and Microsoft) that have already demonstrated interest in supporting CS-education efforts such as improving diversity. All of these stakeholders can be recruited to help in the efforts to grow CER.

Learning from other STEM Education efforts

A number of our recommendations are motivated from the successes and shortcomings of previous STEM education research efforts. From 1990-2005, the National Science Foundation funded the Engineering Coalitions effort [20]. While valuable work was done, it is accepted within the engineering education community [4] that the Coalitions failed to have significant impact. Relevant work was not made widely visible to instructors and "mainstream" researchers beyond the engineering education community and the difficulty of cultural changes was underestimated and lacked a clear reward structure.

The majority of the STEM disciplines, in particular math, chemistry, physics, and biology have education faculty in the disciplinary departments. This includes many top ranked departments that do education research which has impacted teaching and research in the domains. For example, highly ranked physics departments (such as Harvard and Stanford) have physics education research faculty. And, 87% of Physics faculty report familiarity with at least one of 24 evidence-based teaching strategies [5] (the Physics Teacher, a magazine featuring physics education articles directed to teaching practitioners, has a circulation of 8600 [21]). This is in contrast to CS, where most CS faculty could not even name one evidence-based CS practice. There is no fundamental reason preventing CER from emulating the success of physics education research.

In comparison with other STEM fields, CS has a home-field advantage since the people building tools are "in the same tribe" as those doing research and teaching. This puts CS departments in a unique position to focus and channel the world's CS talent in a way that is vastly more productive and evidence-based than it is now. In other words, compared to other fields we have a potentially smoother and quicker path to impact.

## 5. Conclusion

Scalable, evidence-based computing education research addresses both the explosion of demand for high-quality computing education and the diversity of students generating that demand. Leading computer science departments have the intellectual standing, and arguably the moral imperative, to lead in this area; but they need to be open to embracing structural changes in their research and teaching culture and to explore novel collaborations with other departments. Public visibility, funding support, and student demand make it the right time to seize the moment. We urge computer science schools and departments to rise to the challenge.

**Acknowledgements**
Thanks to Alex Aiken, Ann Drobnis, Matt Dwyer, Michael Ernst, Ed Fox, Mark Guzdial, Andre van de Hoek, Hank Levy, Ran Libeskind-Hadas, Daniel Lopresti, Lori Pollack, Leo Porter, Debra Richardson, Anthony Robins, Susan Rodger, Mehran Sahami, and Cliff Shaffer, who commented on early drafts of this document.

This material is based upon work supported by the National Science Foundation under Grant No. (1136993). Any opinions, findings, and conclusions or recommendations expressed in this material are those of the author(s) and do not necessarily reflect the views of the National Science Foundation.